\documentclass[numbers, super,preprint,review,12pt]{elsarticle}

\usepackage{textgreek} 

\usepackage[version=3]{mhchem} 
\usepackage{amsmath} 
\usepackage{mathrsfs}
\usepackage{amssymb} 
\usepackage{amsfonts} 
\usepackage{bbold} 
\usepackage{dsfont} 
\usepackage{stmaryrd} 
\usepackage{graphicx} 
\graphicspath{ {./images/} } 
\usepackage{xcolor, subcaption}
\usepackage{enumitem} 
\usepackage{hyperref} 
\usepackage{tikz}
\DeclareUnicodeCharacter{2212}{\textendash}

\journal{Journal of Catalysis}

\begin{document}

\begin{frontmatter}



\title{Unbiased molecular dynamics for the direct determination of catalytic reaction times : paving the way beyond transition state theory} 


\author[IFPEN Solaize]{Thomas Pigeon\corref{cor1}} 
\ead{thomas.pigeon@ifpen.fr}
\author[IFPEN Solaize]{Manuel Corral Valero}
\author[IFPEN Solaize]{Pascal Raybaud\corref{cor2}}
\ead{raybaud@ifpen.fr}

\affiliation[IFPEN Solaize]{organization={IFP Energies Nouvelles},
            addressline={Rond-Point de l’Echangeur de Solaize, BP 3}, 
            city={Solaize},
            postcode={69360}, 
            state={},
            country={France}}


\begin{graphicalabstract}
\includegraphics[width=14cm]{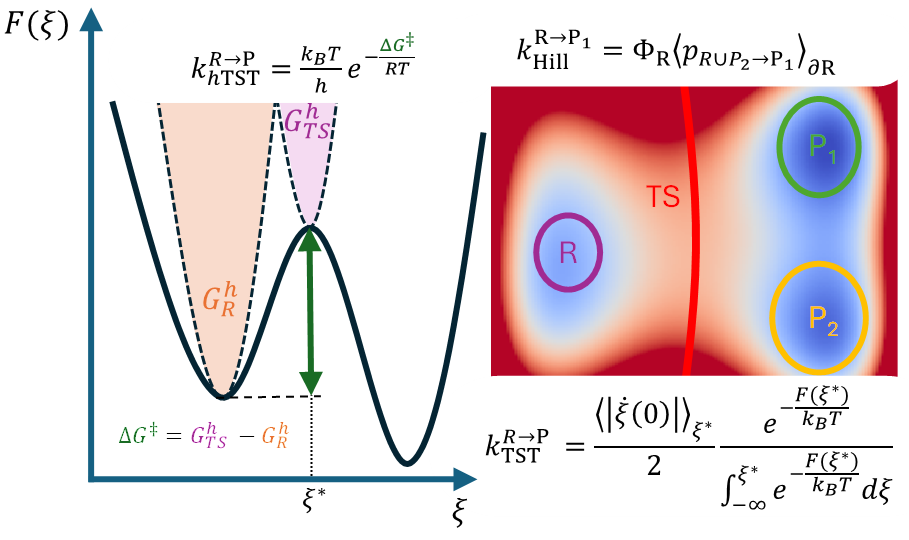}
\end{graphicalabstract}



\end{frontmatter}



\section{Introduction}
\label{sec:intro}

Industrial catalytic processes such as Fischer-Tropsch\cite{Corral_Valero2013} synthesis or biomass\cite{Li2016} conversion  involve intricate reaction mechanisms where multiple pathways compete and formed intermediates can undergo side reactions. The accurate prediction of reaction rate constants is fundamental to understand and optimize catalytic materials as well as catalytic processes. In heterogeneous catalysis, elementary steps such as molecular adsorption, surface diffusion, bond activation, and product desorption govern the overall kinetics and selectivity of complex reaction networks. The ability to compute these rate constants from first principles approaches, such as density functional theory (DFT), provides crucial insights for rational catalyst design, enabling the identification of optimal chemical formulation,\cite{norskov_origin_2002}, active site geometries, \cite{van_santen_complementary_2009} support effects,\cite{Shahrokhi2025} and reaction conditions.

Nevertheless, numerous theoretical challenges still exist in the way to calculate accurately the rate constant. After analyzing some earlier fundamental concepts based on Transition State Theory (TST), this perspective aims at discussing alternative ways for direct determination of rate constant through unbiased molecular dynamics.

\paragraph{Theoretical Framework for Rate Constant Calculations} The most widely employed theoretical approach for computing reaction rate constants is Transition State Theory (TST), originally developed by Eyring, Evans, and Polanyi.\cite{EyringPolanyi1931, Eyring1935, Evans1935} 

The essence of TST lies in the partitioning of configuration space $\Omega = \mathcal{V}^{N}$, with $N$ the number of particles and $\mathcal{V}$ the simulation volume, into two distinct complementary regions $\mathscr{R}$ and $\mathscr{P}$ associated with reactants and products such that $\mathscr{R} \cup \mathscr{P} = \Omega$, separated by a boundary known as the transition state (TS). It is crucial to emphasize that the TS does not correspond to a single configuration but rather represents an ensemble of configurations. Considering a simplified scenario involving a single transition pathway between reactants $\mathscr{R}$ and products $\mathscr{P}$, a widely adopted heuristic for defining the transition state surface involves identifying it as the surface encompassing the highest first-order saddle point encountered along the minimum energy path (MEP) connecting the reactants' and products' basins, and perpendicular to the MEP at the saddle point. This explains the frequent misconception of the TS as a singular configuration, specifically the aforementioned first-order saddle point. Many methodologies such as Nudged Elastic Band (NEB)\cite{Henkelman2000} or the string method\cite{E2002} aim at identifying minimum energy paths and then locate this first order saddle-point in order to estimate the activation energy and the associated rate, generally using the harmonic approximation.


A simplification of the general TST framework arises when employing the harmonic approximation around the transition state and reactant state, leading to harmonic Transition State Theory (hTST). In this case, the rate constant reduces to the well-known Eyring--Polanyi expression:\cite{EyringPolanyi1931, Eyring1935, Evans1935}
\begin{equation}
\label{eq:Eyring_polanyi}
k^{\text{hTST}}_{R\rightarrow P} = \frac{k_\mathrm{B} T}{h} e^{-\frac{\Delta G^{\ddagger}}{RT}},
\end{equation}
where $h$ is Planck's constant and $\Delta G^{\ddagger}$ is the activation free energy computed as the difference between the harmonic approximation of the transition state and reactant free energies. 

General TST calculations including in $k^{\text{TST}}_{R\rightarrow P}$ anharmonicity and configurational entropy offer improved accuracy by incorporating the complete free energy profile along the reaction coordinate, but require extensive sampling of the Boltzmann-Gibbs distribution through enhanced sampling techniques like blue-moon sampling\cite{CARTER1989}, metadynamics,\cite{Laio2002} adaptive biasing force,\cite{Darve2001} or slow-growth,\cite{Woo1997, Jarzynski1997} just to cite a few. 

Finally, TST necessarily overestimates reaction rates due to the assumption that trajectories crossing the transition state surface commit irreversibly to products. In catalytic systems, this limitation is particularly pronounced for reactions involving flexible adsorbates, multiple transition channels, or significant solvent effects.\cite{PETERS2017} This is addressed by introducing a transmission coefficient $\kappa \in (0,1)$:
\begin{equation}
\label{eq:transmission_coef_general}
k_{R\rightarrow P} = \kappa k^{\text{TST}}_{R\rightarrow P}.
\end{equation}
The transmission coefficient accounts for dynamical effects, and quantum effect can be included if the underlying dynamics address them (see for example the ring polymer MD formalism\cite{Manolopoulos2004}). In a classical setting, the dynamical effects include the transition surface recrossings. Its accurate determination requires molecular dynamics simulations which represents an additional computational cost.\cite{Chandler1978,Bennett1977}

The computational cost hierarchy among these methods reflects a fundamental trade-off between accuracy and efficiency. Harmonic TST represents the most computationally efficient approach, requiring only geometry optimization and harmonic frequency calculations for the reactant and transition state configurations. This method is used in countless \textit{ab-initio} studies of the reactivity in enzyme catalysis,\cite{warshel_computer_1976,kamerlin_at_2013} homogeneous\cite{Sciortino2023} and heterogeneous\cite{Zijlstra2020} catalysis but relies on strong approximations that may fail for systems with significant anharmonicity and entropy effects.\cite{Collinge2020} Recent applications to heterogeneous catalysis of the general framework incorporating the complete free energy profile have demonstrated the importance of free energy corrections for surface reactions.\cite{Collinge2020, Bonati2023, Gešvandtnerová2024} These approaches involve a drastic increase of the computational cost due to the required sampling of Boltzmann-Gibbs measure. Finally, in some cases, important recrossing or post transition state bifurcation may lead to failures of TST as recently illustrated in Refs.~\citenum{Gešvandtnerová2024, Gešvandtnerová2025}, for the dehydration of isobutanol, one relevant case study that we will highlight in this perspective.

\paragraph{The Hill Relation: An Alternative Framework} 
An alternative approach to computing reaction rates is provided by the Hill relation:\cite{Hill2012}
\begin{equation}
\label{eq:hill_relation}
k^{\text{Hill}} = \Phi_R \langle p_{R\rightarrow P} \rangle_{\partial R},
\end{equation}
where $\Phi_R$ is the flux of dynamic trajectories leaving the reactant state R, and $\langle p_{R\rightarrow P} \rangle_{\partial R}$ is the committor probability (the probability of reaching products before returning to reactants)\cite{Hummer2004, Vanden-Eijnden2006} averaged over exit distribution of $R$ defined on the boundary of the reactant state $\partial R$. This formulation thus incorporates dynamical effects intrinsically and provide rate constants which are exact for the underlying stochastic dynamics.\cite{BAUDEL2023,Lelievre2024}

The averaged committor probability can be estimated using advanced rare event sampling methods such as Transition Interface Sampling (TIS),\cite{vanErp2003} Forward Flux Sampling (FFS),\cite{allen2005} or Adaptive Multilevel Splitting (AMS).\cite{Cerou2007} These methods employ splitting estimators that express the overall transition probability as a product of conditional probabilities across multiple interfaces between reactants and products. While computationally demanding, these approaches have shown success in various types of rare event studies, see Refs.~\citenum{Hussain2020, Cabriolu2017, Reiner2023, Riccardi2019} for some applications. Even-though some studies of catalytic reactions were done\cite{Bučko2009, Bučko2011, Schwartz2022, Gešvandtnerová2024} using Transition Path Sampling (TPS)\cite{Dellago2003} -- the basis of TIS -- the calculation of rates using the Hill relation is still not common in the catalysis field. 

\paragraph{Computational Considerations and the Machine Learning methods} Methods based on the Hill relation represent the most computationally demanding approach, as they require extensive molecular dynamics simulations to estimate committor probabilities through rare event sampling methodologies. The computational burden scales with the rarity of the reactive event, limiting applications to relatively simple model systems.

However, the recent emergence of machine learning interatomic potentials (MLIPs) for ground\cite{Behler2017, Deringer2019, Mueller2020, Unke2021, Manzhos2021} and excited states\cite{Westermayr2021} has fundamentally altered this computational landscape. Modern MLIPs, trained on high-level quantum mechanical data, achieve near-DFT accuracy while providing computational efficiency that can be, if not comparable to classical force fields, a drastic acceleration measurable in multiple orders of magnitude.\cite{Batatia2022mace, Batatia2022Design, Drautz2019,JACOBS2025} Moreover, in recent years, various foundational models have been developed and are now readily available.\cite{JACOBS2025, batatia2023foundation, Bochkarev2025} Fine-tuning these models to specific datasets enables the achievement of satisfactory precision at a really affordable training cost. This breakthrough enables the extensive molecular dynamics simulations required for committor probability estimation to become computationally feasible for complex catalytic systems.

The accessibility of Hill relation-based methods through MLIPs opens transformative possibilities for heterogeneous catalysis. These methods can now tackle problems including reactions with complex multi-states mechanisms, surface diffusion processes, and long-timescale phenomena that govern selectivity in catalytic networks. The ability to simulate large-scale systems over extended time periods while maintaining close to quantum-level accuracy enables the study of realistic catalyst models with defects, promoters, and support interactions\cite{JACOBS2025}. Recent applications have demonstrated the potential of MLIP-enhanced rare event sampling in materials science and are beginning to emerge in catalysis research\cite{Bocus2023, Bonati2023, Tang2024, Bocus2025}.

This paradigm shift motivates the present viewpoint, which explores how MLIP-enhanced molecular dynamics can make previously prohibitive computational approaches accessible for practical catalyst studies. This perspective will firstly introduce some fundamental aspects on the Hill relation framework in close connection with the Adaptive Multilevel Splitting (AMS) approach (Section~\ref{sec:Hill}). Then, we will underline some methods to identify collective variables and reaction coordinates (Section~\ref{sec:CVs}). We finally highlight the potentiality of this method in conjunction with MLIPs on two case studies relevant for catalysis : water activation on an alumina surface and dehydration of protonated isobutanol in gas phase (Section~\ref{sec:illustration}).

\section{Computing reaction rates with the Hill relation}
\label{sec:Hill}

\subsection{Flux formulation}

We consider a system of $N$ atoms with configurations $\textbf{q} \in \Omega = \mathcal{V}^N$ (as defined in Introduction) and momenta $\textbf{p}\in \mathbb{R}^{3N}$. The metastable states can be defined using dedicated collective variables $\zeta_R$ and $\zeta_P$. To estimate the flux $\Phi_R$, one introduces a thin boundary layer around it to ensure sampling of trajectories "truly" leaving $R$. For instance, if the reactant state $R$ is defined in terms of a one dimensional collective variable $\zeta_R$ as
\[
R = \{ \textbf{q} \in \Omega \;|\; \zeta_R(\textbf{q}) \leqslant z_R \},
\]
then $\partial R = \{ \textbf{q} \;|\; \zeta_R(\textbf{q}) = z_R \}$ is replaced in practice by a narrow shell 
\[
\partial R_\varepsilon = \{ \textbf{q} \;|\; z_R \leq \zeta_R(\textbf{q}) \leq z_R + \varepsilon \},
\]
with $\varepsilon$ is small enough so that configurations in $\partial R_\varepsilon$ still represent genuine exits from $R$ but large enough to ensure that the excursion of out $R$ do not have a negligible length. In this setting, the actual excursions out of $R$ corresponds to the configurations such that $\zeta_R(\textbf{q}) \geqslant z_R + \varepsilon$. 

The flux is estimated using a trajectory starting in the $R$ state (or basin) and crossing its boundary back and forth many time, as the inverse of the average loop time:
\begin{equation}
    \label{eq:loop_time}
    \widehat{\Phi}_R = \left(\frac{1}{n_\mathrm{loop}}\sum_{i=1}^{n_\mathrm{loop}} t_\mathrm{loop}^i \right)^{-1}.
\end{equation}
A loop is a portion of trajectory starting when the trajectory enters $R$ ($\zeta_R(\textbf{q}) \leqslant z_R$) and finishing the next time it returns to basin $R$ after having done at least one significant excursion ($\zeta_R(\textbf{q}) \geqslant z_R + \varepsilon$), see Figure~\ref{fig:loops_and_splitting_final_a}. Simultaneously, $t_\mathrm{loop}^i$ represents the time spent in each loop.

For each loop, the first configuration such that $\zeta_R(\textbf{q}) \geqslant z_R + \varepsilon$ is saved, as such configurations sampling the exit distribution on $\partial R$ are later used to initialize the AMS algorithm, allowing to estimate the averaged committor function against this exit distribution in Equation~\eqref{eq:hill_relation}.   

\subsection{Splitting estimators for committor evaluation}
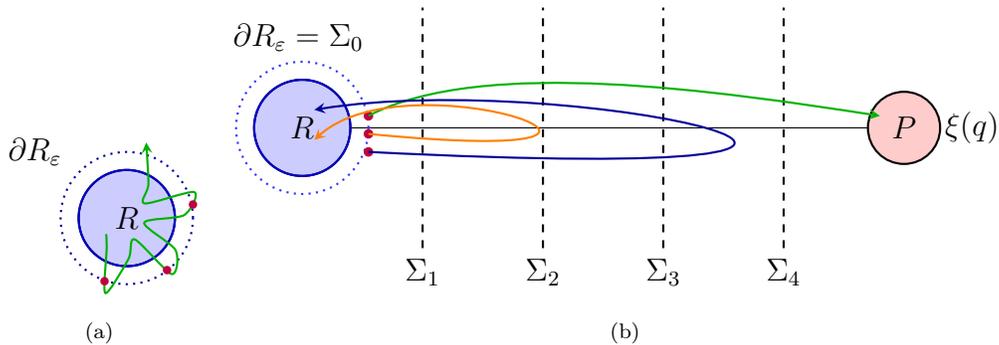
\begin{figure}[ht]
    \centering
    \begin{subfigure}[t]{0.2\textwidth}
    \centering
    \begin{tikzpicture}[scale=0.8,>=stealth]

        \coordinate (Rc) at (0,0);
        \def\R{0.8}
        \def\Reps{1.1}

        \begin{scope}
            \clip (Rc) circle (\Reps);
            \fill[blue!10] (Rc) circle (\R);
        \end{scope}

        \draw[thick,dotted,blue!60!black] (Rc) circle (\Reps);
        \node[above left] at (-0.9,0.7) {$\partial R_\varepsilon$};

        \draw[fill=blue!20, thick] (Rc) circle (\R);
        \draw[thick,blue!80!black] (Rc) circle (\R);
        \node at (Rc) {$R$};

       \draw[thick,->,green!70!black,smooth]
    plot[smooth] coordinates {
      (-0.344,-0.267)
      (-0.333,-0.493)
      (-0.355,-1.248)
      (0.027,-0.682)
      (0.132,-0.355)
      (0.743,-0.931)
      (0.809,-0.531)
      (0.290,-0.078)
      (0.954,0.050)
      (1.118,0.433)
      (0.298,0.244)
      (0.360,0.565)
      (0.322,1.258)
       };
        \fill[purple] (-0.37,-1.05) circle (0.07);
        \fill[purple] (0.669,-0.865) circle (0.07);
        \fill[purple] (1.1,0.226) circle (0.07);


    \end{tikzpicture}
    \caption{}
    \label{fig:loops_and_splitting_final_a}
    \end{subfigure}
    \hfill
    \begin{subfigure}[t]{0.78\textwidth}
    \centering
    \begin{tikzpicture}[scale=0.8,>=stealth]

        \draw[->] (-0.5,0) -- (10.5,0) node[right] {$\xi(q)$};

        \coordinate (Rc) at (0,0);
        \def\R{0.8}
        \def\Reps{1.1}

        \begin{scope}
            \clip (Rc) circle (\Reps);
            \fill[blue!10] (Rc) circle (\R);
        \end{scope}

        \draw[thick,dotted,blue!80] (Rc) circle (\Reps);
        \node[above left] at (1.2,1.1) {$\partial R_\varepsilon = \Sigma_0$};

        \draw[fill=blue!20, thick] (Rc) circle (\R);
        \draw[thick,blue!80!black] (Rc) circle (\R);
        \node at (Rc) {$R$};

        \draw[fill=red!20, thick] (10,0) circle (0.6);
        \node at (10,0) {$P$};

        \foreach \x/\label in {2/$\Sigma_1$,4/$\Sigma_2$,6/$\Sigma_3$,8/$\Sigma_4$} {
            \draw[thick,dashed] (\x,-2.0) -- (\x,2.0);
            \node[below] at (\x,-2.0) {\label};
        }

        \filldraw[purple] (\Reps,0.2) circle (2pt);
        \draw[->,thick,green!70!black,smooth]
            (\Reps,0.2) .. controls (3,1.2) and (7,0.6) .. (9.6,0.2);

        \filldraw[purple] (\Reps,-0.1) circle (2pt);
        \draw[->,thick,orange,smooth]
            (\Reps,-0.1) .. controls (7.2,-0.7) and (1.8,1.25) .. (0.2,-0.2);

        \filldraw[purple] (\Reps,-0.4) circle (2pt);
        \draw[->,thick,blue!60!black,smooth]
            (\Reps,-0.4) .. controls (13,-1.0) and (4.5,1.) .. (0.2,0.3);

        
    \end{tikzpicture}
    \caption{}
    \label{fig:loops_and_splitting_final_b}
    \end{subfigure}

    \caption{
    Schematic representation of 
    \textbf{(a)}~a loop trajectory starting in the reactant basin $R$, exiting and re-entering multiple times. The purple dot marks the first crossing of $\partial R_\varepsilon$ after leaving $R$ and
    \textbf{(b)}~splitting estimators with interfaces $\Sigma_i$ taken as iso-levels of a reaction coordinate $\xi$. The reactant basin $R$ and product basin $P$ are represented as circles. Reactive trajectories (green) cross all interfaces and end in $P$, while non-reactive ones (orange, blue) return to $R$ after partial progress (the orange trajectory crosses $\Sigma_1$ but not $\Sigma_2$).
    }
    \label{fig:loops_and_splitting_final}
\end{figure}

Direct evaluation of $\langle p_{R\to P}\rangle_{\partial R}$ by brute-force simulation is impractical since $p_{R\to P}$ is typically very small in rare-event settings. Splitting methods introduce a sequence of interfaces
\[
\Sigma_0 = \partial R_\varepsilon, \, \Sigma_1, \, \dots, \, \Sigma_M = \partial P,
\]
such that any reactive trajectory must cross all interfaces in order (see Figure~\ref{fig:loops_and_splitting_final_b}). The boundary-averaged probability can then be written as a product of conditional probabilities,
\begin{equation}
\label{eq:prob_product}
p_{R\to P}(\partial R_\varepsilon) \;=\; p_{R\to \Sigma_1}(\partial R_\varepsilon) \,\prod_{j=1}^{M-1} p_{\Sigma_j \to \Sigma_{j+1}}(\Sigma_j)\; \times\; p_{\Sigma_M \to P}(\Sigma_M),
\end{equation}
where $p_{\Sigma_j \to \Sigma_{j+1}}(\Sigma_j)$ denotes the probability to hit $\Sigma_{j+1}$ before returning to $R$, conditional on first hitting $\Sigma_j$.

Methods like TIS,\cite{vanErp2003} FFS\citenum{allen2005} or AMS,\cite{Cerou2007,Lopes2019} based on such a splitting estimator, are designed to obtain estimates of the conditional probabilities $p_{\Sigma_j \to \Sigma_{j+1}}(\Sigma_j)$. Here, we will focus on the description of the AMS method. It has the advantage of automatically placing the surfaces $\Sigma_j$ so that all conditional probabilities $p_{\Sigma_j \to \Sigma_{j+1}}(\Sigma_j)$ are equal, which is a required feature as it minimize the variance of the splitting estimator. 

\subsection{Adaptive Multilevel Splitting (AMS)}
The algorithm starts with $N_{\mathrm{rep}}$ replicas from $\partial R$ (exit set), propagate until each replica hits either $R$ or $P$, and define a progress score with a one dimensional reaction coordinate  $\xi$. At iteration $j$, the algorithm kills the fraction $\eta_j^{\mathrm{killed}}/N_{\mathrm{rep}}$ with the $k^\mathrm{th}$ worst score $\xi_\mathrm{min}$ and replace them by clones of one of the remaining ones branched at $\xi_\mathrm{min}$. The process continues until the dynamics reaches $R$ or $P$. This procedure is further detailed in SI Section~1 and illustrated in Figure~S.1. The splitting estimator of the boundary-averaged committor is
\begin{equation}
\label{eq:AMS_proba_estimator}
\widehat{p}_{\Sigma_j\to\Sigma_{j+1}}\left( \Sigma_j \right)
= 1-\frac{\eta_j^{\mathrm{killed}}}{N_{\mathrm{rep}}},
\qquad
\widehat{k}^{\mathrm{AMS}}
=\widehat{\Phi}_R\;\prod_{j=1}^{n_{\max}}\!\left(1-\frac{\eta_j^{\mathrm{killed}}}{N_{\mathrm{rep}}}\right),
\end{equation}
where $n_{\max}$ is the number of iterations of AMS algorithm done until all the replicas end in $P$. AMS places the “interfaces” adaptively, and assuming that the fraction with the worst score is constant, meaning that there are always $k$ replicas killed at each iteration, the conditional probabilities are thus equalized, which reduces the variance of the estimator.

By repeating the AMS estimation $M_\mathrm{real}$ times, the empirical variance on the AMS probability estimator can be estimated. Combining it to the variance of the estimator~\eqref{eq:loop_time} of $\widehat{\Phi}_R$, the confidence intervals, or errors, of the estimated rate can be computed as detailed in Ref.~\citenum{Pigeon2023}. The width of these confidence intervals depends on:
\begin{itemize}
    \item the reaction coordinate $\xi$;
    \item the number of replicas $N_\mathrm{rep}$;
    \item the number of realizations of AMS $M_\mathrm{real}$;
    \item the number of loops $n_\mathrm{loop}$.
\end{itemize}
In practice, as sampling loops is fast, $n_\mathrm{loop}$ is large enough and the main contribution on the rate errors comes from the other parameters, as discussed in the section dedicated to the examples. 

Theoretically, any AMS realizations yield an estimation of the committor averaged against the exit distribution on $\partial R$, though in practice both the definition of metastable states using the CVs $\zeta_R$,$\zeta_P$ and the choice of reaction coordinate $\xi$ impact heavily the computational cost, and hence the feasibility, of the calculation. This is explained next section.

\section{On the choice of collective variables and reaction coordinates}
\label{sec:CVs}

\subsection{Choice of collective variable to estimate reaction rates}

A central difficulty in applying both TST and Hill-based approaches to complex catalytic systems lies in the choice of collective variables (CVs). These variables can serve distinct purposes: defining metastable states ($R$ and $P$), organizing the sampling of transition pathways or defining the transition state surface.

\paragraph{Reaction coordinates in TST}  
In TST, the rate constant is obtained by counting crossings of a dividing surface $\Sigma(z_\mathrm{TS})$ defined as the iso-surface of a one-dimensional reaction coordinate $\xi$.  The choice of $\xi$ directly controls the accuracy of TST: a poor coordinate typically yields a very small transmission coefficient $\kappa$, reflecting frequent recrossings of $\Sigma(z_\mathrm{TS})$. In the extreme case where $\xi$ is essentially uncorrelated with the true reaction progress, the computed TST rate may be orders of magnitude too large, and correcting it via $\kappa$ becomes statistically impractical. Thus, for TST-based approaches, the reaction coordinate is not merely a tool for efficiency but a central source of systematic bias. The optimal choice of reaction coordinate is the committor function.\cite{VandenEijnden2005}

\paragraph{Defining metastable states in Hill-based methods}  
When using the Hill relation, the reactant and product basins must be specified unambiguously to ensure that sampled trajectories correspond to the desired event. In practice, this can be achieved through indicator functions based on dedicated (or not) descriptors $\zeta_R$ and $\zeta_P$. Suitable choices include distances to reference structures, coordination numbers, local order parameters, or environment-sensitive fingerprints tailored to catalytic active sites. The key requirement is robustness: $\zeta_R$ and $\zeta_P$ must clearly discriminate configurations belonging to $R$ and $P$ from any other potentially encountered configuration along the reaction path, so that no undesired events are sampled. 

\paragraph{Reaction coordinates in Hill-based methods}  
The progress variable $\xi$ is used to define a sequence of interfaces $\Sigma_j$ between $R$ and $P$ to ease the estimation of the committor probability. The rate estimate is unbiased regardless of the choice of $\xi$, provided interfaces are ordered such that all reactive trajectories cross them sequentially. Here, the impact of $\xi$ is purely statistical: a poor choice (i.e.: intrinsically different from the committor function) increases the variance of the splitting estimators, raising the computational cost but not biasing the result. The committor probability itself is generally inaccessible, or require iterative sampling-training procedure using neural networks such as in Refs~\citenum{pmlr-v145-rotskoff22a, Kang2024, wang2025}

\subsection{Automated construction of CVs.}  
\label{sec:CV_learning}
Defining good collective variables remains one of the central challenges in rare-event simulations. Beyond physical intuition, several systematic approaches have been developed to construct CVs with a clear mathematical foundation.

\paragraph{Unsupervised methods} A first family is based on the variance based dimensionality-reduction methods, such as Principal Components Analysis (PCA) and AutoEncoders (AEs), which identify directions or nonlinear embeddings associated with the largest variance of the configurational distribution. While not necessarily optimal dynamically, they often correlate with meaningful collective motions, like minimum energy paths when trained on Boltzmann-Gibbs distributed data\cite{Lelièvre_2024}. Their success depends on dominant structural fluctuations participating in activated processes and the availability of configurations for training. The second family is based on spectral approaches approximating the first eigenfunctions of dynamical operators, such as the dynamics infinitesimal generator $\mathcal{L}$ or the time propagator $\mathscr{P}_t$ representing slow dynamical modes and metastable states. Time-lagged independent component analysis (TICA)\cite{Molgedey1994, Naritomi2011} identifies linear combinations of descriptors for the leading eigenfunctions, while state-free reversible VAMPNets\cite{Chen2019} and DeepTICA\cite{Bonati2021} extend this to nonlinear functions using neural networks which is similar to extended dynamical mode decomposition\cite{Williams2015} and variational conformational kinetics\cite{Nuske2014}. Kernel methods presenting an alternative to neural networks and can also be used to approximate these first eigenfunctions of $\mathcal{L}$\cite{KLUS2020, Nateghi2025}. Both spectral and variance based methods require a good sampling of the Boltzmann-Gibbs measure  which can be alleviated by using iterative methods, such as those proposed in Refs.~\citenum{chen2018, Belkacemi2021}. 

\paragraph{Supervised methods} They are based on an expert knowledge to label configurations. Employing classifiers like support vector machine (SVMs) or Fisher discriminants analysis to define CVs proved their efficiency\cite{Murphy2022, Sultan2018, Bonati2020}. Alternatively, path-based variables compare configurations to references paths using similarity measures \cite{Branduardi2007}. It can be coupled with local environment classifiers\cite{Rogal2019} or neural networks methods\cite{Fröhlking2024} to obtain even better performances. 

\paragraph{Semi-supervised method} Lastly, committor functions $p_{R\to P}$ are intrinsically the optimal reaction coordinates. They can be termed semi-supervised in the sense that only the knowledge of $R$ and $P$ is required. Committor function learning methods can be broadly classified into two categories: local regression\cite{Frassek2021, Lopes2019, Jung2023} or physics informed learning techniques\cite{Li2019, yuan2023, Khoo2018, Li2019, rotskoff2020learning, pmlr-v145-rotskoff22a, yuan2023, Kang2024, wang2025, STRAHAN2023, Strahan_2023, Mitchell2024, Li2021, Roux2021, Roux2022, He2022, Chen2023}. Depending on the choice of method, various types of data are required but all of them are generally quite intensive and some iterative sampling--training procedures are proposed to solve this issue\cite{pmlr-v145-rotskoff22a, Jung2023, Kang2024, wang2025}

\section{Two prospective examples}
\label{sec:illustration}

In this section, we illustrate on two relevant case studies how to solve the challenges underlined in previous sections and estimate reaction rates by using the Hill based approach thanks to AMS and molecular dynamics using fine-tuned MACE-MPA-0 MLIP.\cite{batatia2023foundation} The first test case quantifies the rate constant for the formation of a water molecule from two hydroxyls catalyzed by the (100) $\gamma$-alumina surface by using SVM to construct the CV of the reaction. The second one illustrates how to analyze a complex reaction mechanisms in gas phase such as the dehydration of protonated iso-butanol with a CV constructed by a combination of an unsupervised AE method and path-based CVs.

\subsection{\ce{H2O} on the (100) surface of $\gamma$-alumina}
\label{subsec:h2o_alumina}

\begin{figure}[!ht]
  \centering
         \includegraphics[width=\textwidth]{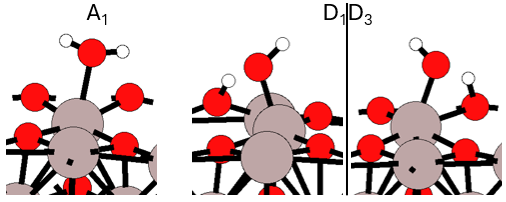}
 Adsorbed water \caption{$A_1$ (left) and dissociated water $D_1D_3$ (right) structures. Color legend: Grey: Al, Red: O, White: H.}
  \label{fig:figure_h2o_al2o3}
\end{figure}

The first prospective study is inspired by a previous recent study \cite{Pigeon2023}, which evaluated various rate constants via the Hill-AMS approach based on ab initio molecular dynamics (AIMD). These rate constants are involved in the reaction network of water adsorbed on the by (100) $\gamma$-alumina surface connecting 5 metastable states: two of them corresponds to the dissociated water molecule (two hydroxyls) while the three others are conformers of the adsorbed water. In what follows, we focus on the transition from the water dissociated states (so called $D_1D_3$ in Ref.\citenum{Pigeon2023}) to water adsorbed ($A_1$ in Ref.\citenum{Pigeon2023}) as illustrated in Figure~\ref{fig:figure_h2o_al2o3} 
Focusing on the $D_1D_3 \to A_1$ transition, we aim at benchmarking:
\begin{itemize}
    \item the consistency of reaction rate constant estimated via AMS using DFT and the finetuned MACE-MPA-0 model;
    \item the impact of the Langevin dynamics friction $\gamma$ and integration time step $\Delta t$ on the estimated rate;
    \item the impact of the number of replicas $N_\mathrm{rep}$ and the number of parallel realizations of AMS $M_\mathrm{real}$ on the estimated error on the estimated rate.
\end{itemize}

In our previous work \cite{Pigeon2023}, the collective variables used to identify numerically whether the system is in a given metastable states were defined in a supervised approach using SOAP\cite{Bartok2013} descriptor, SVM and short trajectory data in the vicinity of each local minimum on the potential energy surface. We refer to this work for further details concerning the construction of these collective variables, the exact same definition of $D_1D_3$ and $A_1$ is used here. In this work, as we only consider the $D_1D_3 \to A_1$ reaction, we used the same SOAP-SVM-$D_1D_3$ CV to define the $D_1D_3$ state and index the progress of trajectories in AMS, in other words, $\zeta_R = \xi$ to use the previously introduced notations. 

The AMS algorithm was implemented using the atomic simulation environment (ASE) python package,\cite{Larsen_2017} and to be able to compare AIMD results to those obtained with finetuned MACE model using the same Langevin dynamics integrator, we performed again the estimation of rates with DFT using the ASE langevin integrator. We observed a small deviation of results which might originate from the different integrators. 


To fine-tune the MACE-MPA-0 foundation model, we built a training, validation and test set by sampling randomly 50 configurations along unbiased MD trajectories at 200 K in the vicinity of the 5 metastable states of the system. This leads to 3 datasets composed of 250 configurations. Using the finetuning settings (reported in SI Section~2), we obtained a test RMSE of 0.7 meV/atoms for the energy and 28.9 meV/\AA~for the forces.  

In SI Table~S.1, we report the estimated $D_1D_3 \to A_1$ reaction rate constants along with the 95\% confidence interval. These confidence intervals, estimated using DFT and the fine-tuned MACE-MPA-0 models, overlap, indicating consistent predictions. In other words, considering the same set of parameters (AMS, $\gamma$ and time step size $\Delta t$ of the Langevin dynamics) both the DFT and the fine-tuned MLIP model yield comparable results : $(13.3 \pm 6.3) \times 10^{10}$ $\mathrm{s}^{-1}$ and $(8.93 \pm 4.49) \times 10^{10}$ $\mathrm{s}^{-1}$, respectively. Moreover, variations in the friction parameter $\gamma$ of the Langevin dynamics appear to have little impact on the estimated rate (see entries 2, 3, 12 and 13 in SI Table~S.1), and a similar observation can be made regarding the discretization time step size $\Delta t$ (see entries 4, 10 and 11 in SI Table~S.1). From the relative errors reported in SI Table~S.1, decreasing the time step size (keeping all other parameters fixed) leads to smaller errors, albeit at the cost of increased computational expense. Specifically, reducing $\Delta t$ by a factor of 4 results in a fourfold increase of computational cost. Conversely, larger friction values ($\gamma \sim 5~\mathrm{ps}^{-1}$) seem to produce slightly more accurate results at the same computational cost. This can be rationalized by the fact that, similar to FFS, the AMS algorithm relies on the stochastic nature of the dynamics, where branched trajectories diverge from the reference trajectory. Higher friction enhances the diversity of sampled trajectories through this branching process. Assuming there exists a range of friction values over which the reaction rates of the underlying dynamics remain constant---an assumption supported by the results in SI Table~S.1---the most effective way to accurately estimate these rates is by selecting a friction value in the upper part of this range.

Increasing either the number of AMS evaluations $M_\mathrm{real}$ or the number of replicas $N_\mathrm{rep}$ leads to more accurate predictions. In this particular case, it seems that increasing $N_\mathrm{rep}$ has a more significant effect. This is an expected result, as the central limit theorem for the AMS estimator holds in the limit $N_\mathrm{rep} \to \infty$.\cite{Brehier2015} In practice, the finite values of $N_\mathrm{rep}$ required to reach this asymptotic regime most likely depend on the dynamics, specifically on the friction parameter, the temperature, and the scarceness of the considered event. We could expect the following tendency: the smaller the rates, friction, and temperature, the larger the required $N_\mathrm{rep}$.

Considering the various sets of parameters for AMS and Langevin dynamics, the rate constant is estimated to be comprised between $8 \times 10^{10}$ $\mathrm{s}^{-1}$ and $20 \times 10^{10}$ $\mathrm{s}^{-1}$, whereas the hTST rate constant estimated via Equation~\eqref{eq:Eyring_polanyi} for the $D_1D_3 \to A_1$ reaction was about $11 \times 10^{11}$.\cite{Pigeon2023} This 10 times larger value can be explained by the error introduced by several effects : the harmonic approximation, leading to a poor estimation of entropic effects and the recrossings of the TS surface. 

\subsection{Protonated isobutanol (\ce{iBuOH2+}) reactions in gaz phase}
\label{subsec:ibuoh2_gaz}

\begin{figure}[!ht]
  \centering
         \includegraphics[width=\textwidth]{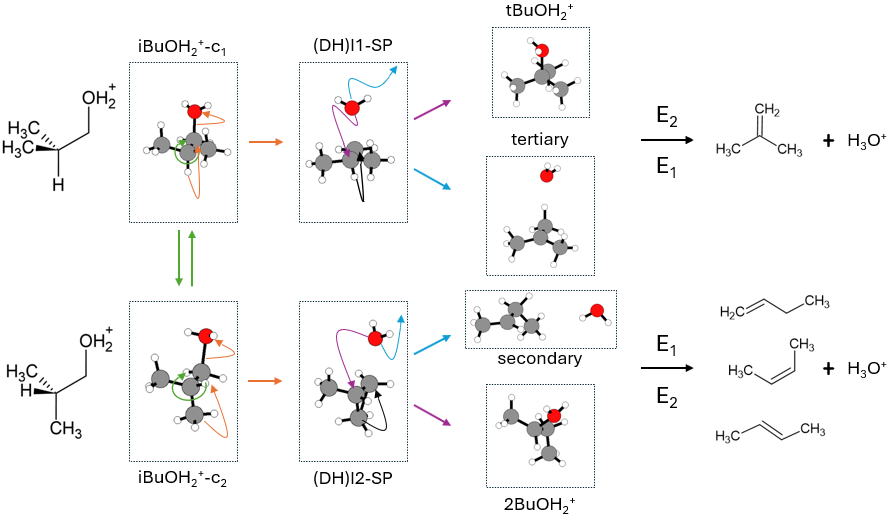}
  \caption{Key intermediates and saddle points (SP) involved in the transformation of two conformers ($c_1$ and $c_2$) of the protonated isobutanol into linear or branched butene in gaz phase. The curved arrows on top of the molecules schematically represent the motion leading to the various transformation observed in MD trajectories. Color legend: Grey: C, Red: O, White: H.}
  \label{fig:figure_iBuOH2}
\end{figure}

The second prospective example concern the dehydration and isomerization of isobutanol which has been previously studied in gas phase and in the chabazite zeolite by using hTST in Ref.~\citenum{Gešvandtnerová2022}, while the significance of dynamical effects was further emphasized in Refs.~\citenum{Gešvandtnerová2024,Gešvandtnerová2025}. The authors applied TST in conjunction with the blue-moon ensemble approach to compute more accurate free energy profiles. They also introduced a simplified gas phase system, protonated iso-butanol (\ce{iBuOH2+}), to investigate dynamical effects by determining the transmission coefficient, $\kappa$. An intriguing result was that the dehydration and isomerization reactions involved in the the transformation share the same transition state. The commitment to the isomerized alcohols or butenes products occurs only after crossing the transition state region. The structures of the corresponding saddle points, along with the metastable states for the simplified system in the gas phase, are presented in Figure~\ref{fig:figure_iBuOH2}. Being inspired from Ref.~\citenum{Gešvandtnerová2024,Gešvandtnerová2025}, the present section analyzes reactive trajectories sampled at 500 K, originating from both the $c_1$ and $c_2$ conformations (see Figure~\ref{fig:figure_iBuOH2}) of the \ce{iBuOH2+} metastable states in gas phase.

The first step is to construct a fine-tuning dataset to obtain an accurate MLIP. To achieve this, we generated four hundred 2 ps-long MD trajectories ($\Delta t = 1 \, \text{fs}$) at 500 K using ASE's Langevin integrator, starting from the saddle point (or harmonic transition state) configurations corresponding to the (DH)I1 and (DH)I2 reactions reported in Ref.~\citenum{Gešvandtnerová2025} (200 trajectories for each configuration). Upon visually inspecting all final points of these trajectories, we observed that, at 500 K, the protonated tert-butyl alcohol \ce{tBuOH2+} was never formed, consistent with the results presented in Ref.~\citenum{Gešvandtnerová2025}. Along each trajectory, a configuration was extracted every 50 fs up to 550 fs, resulting in 12 configurations per trajectory. This final dataset of 12,000 configurations was divided into three subsets for training, validation, and testing. The fine-tuning settings, detailed in the SI Section~2, yielded a test RMSE of 0.6 meV/atom for the energy and 19.7 meV/\AA~for the forces.

In the next step, to construct collective variables for identifying metastable states and tracking progress along a reaction pathway, we employed a combination of path-based collective variables (PCVs),\cite{Branduardi2007} autoencoders,\cite{Lelièvre_2024} and MACE model's atomic environment descriptors.\cite{Batatia2022mace} This approach leverages the detailed information encoded in the MACE descriptors to build meaningful collective variables. Moreover, as we use a finetuned MACE model, the MACE descriptor is computed every MD steps to predict energies and force so using it to compute collective variables does not require any additional computational cost. 

As discussed in Section~\ref{sec:CV_learning}, the PCVs are constructed by computing the similarities between a given feature vector $\textbf{x}$ and those of a predefined path $\left(\textbf{x}_i\right)_{0 \leqslant i \leqslant \ell -1}$ of length $\ell$ which could be a minimum energy path (or intrinsic reaction coordinate IRC). These features composing the vector $\textbf{x}$ could be key distances or angles as in Refs.~\citenum{Gešvandtnerová2024, Gešvandtnerová2025}. In its original formulation,\citenum{Branduardi2007} the progress variable along the path is expressed as:
\begin{equation}
   \label{eq:s_pcv}
   s(\textbf{x}) = \frac{1}{\ell -1}\frac{\displaystyle \sum_{i=0}^{\ell -1} i\,\mathrm{e}^{-\lambda\left|\textbf{x} - \textbf{x}_i\right|^2}}{\displaystyle \sum_{i=0}^{\ell -1} \mathrm{e}^{-\lambda\left|\textbf{x} - \textbf{x}_i\right|^2}},     
\end{equation}
while the variable measuring the distance to the path is given by:
\begin{equation}
   \label{eq:z_pcv}
   z(\textbf{x}) = -\frac{1}{\lambda} \ln \left( \sum_{i=0}^{\ell -1} \mathrm{e}^{-\lambda\left|\textbf{x} - \textbf{x}_i\right|^2}\right).  
\end{equation}
In these equations, Gaussian similarities are calculated by first determining the squared distance between $\textbf{x}$ and $\textbf{x}_i$. The $\lambda$ parameter is a scaling factor that should be of the same order of magnitude of the inverse squared distance between two successive images along the path. A difficulty arises if one wants to replace $\textbf{x}$ with the MACE descriptor as it resides in a 256-dimensional space for the MACE-MPA-0 model. Indeed, in high-dimensional spaces, understand that the vector $\textbf{x}$ is composed of many features, computing distances can be challenging,\cite{Aggarwal2001} and intuitions derived from low-dimensional examples may not apply. For instance, in very high-dimensional spaces, the ratio of the minimum to the maximum Euclidean distance $\textbf{x} - \textbf{x}_i$ approaches one,\cite{Aggarwal2001} leading to less meaningful Gaussian similarities and, consequently, less reliable PCVs. For this reason, such approaches are often coupled with supervised or unsupervised methods to reduce the dimensionality of the input features.\cite{Rogal2019, Fröhlking2024} Additionally, permutation invariance with respect to atom indices must be enforced because their ordering, similar to the ordering of atomic positions, is arbitrary and should not have an impact on the identified CVs. To address these challenges, we employ a modified autoencoder (AE).

Autoencoders consist of two functions, $f_\mathrm{enc}$ and $f_\mathrm{dec}$. The encoder $f_\mathrm{enc}$ maps a high-dimensional space (e.g., $\mathbb{R}^D$) to a lower-dimensional space (e.g., $\mathbb{R}^d$ with $d \ll D$), while the decoder $f_\mathrm{dec}$ performs the reverse mapping. Training an AE involves minimizing the reconstruction error, $\sum_i \left( X_i - f_\mathrm{dec}(f_\mathrm{enc}(X_i))\right)^2$, where $X_i$ are points in the high-dimensional space. This process identifies the optimal functions $f_\mathrm{enc}^*$ and $f_\mathrm{dec}^*$. The optimal encoder $f_\mathrm{enc}^*$ projects the input configurations $X_i$ into the small dimensional features $\textbf{x}_i$ that captures the largest differences between the various $X_i$, enabling their accurate reconstruction by the optimal decoder $f_\mathrm{dec}^*$. 

In our study, we employed an AE architecture to capture the most significant variations in atomic environment descriptors across trajectories initiated at the (DH)I1 and (DH)I2 saddle points. The AE's high-dimensional inputs were the matrices of size $(N_\mathrm{atoms}, N_\mathrm{desc})$ containing the $N_\mathrm{atoms}$ MACE descriptors of size $N_\mathrm{desc}=256$. To ensure permutation invariance with respect to atom indices, we designed a modified AE architecture, detailed in SI Section~3, and the calculation of the permutation-invariant reconstruction error $\left( X_i - f_\mathrm{dec}(f_\mathrm{enc}(X_i))\right)^2$ is described in SI Section~4.

\begin{figure}[!ht]
  \centering
    \begin{subfigure}{0.45\textwidth}
         \centering
         \includegraphics[width=\textwidth]{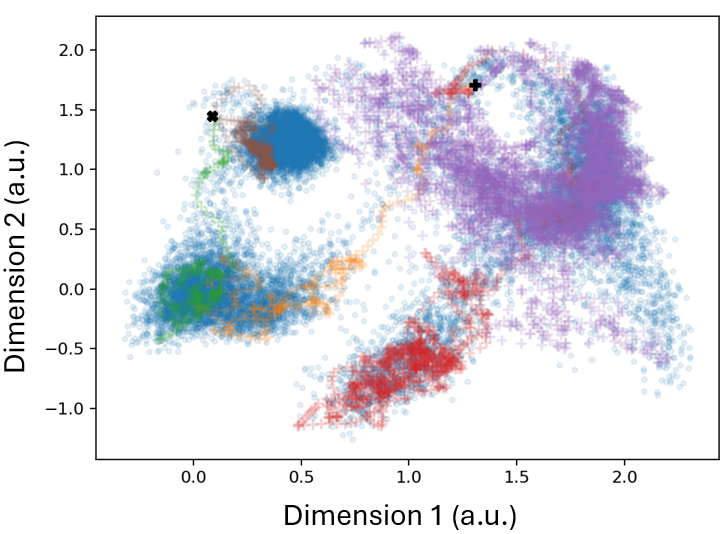}
         \caption{\label{fig:scatter_0_1}}
     \end{subfigure}
     \begin{subfigure}{0.45\textwidth}
         \centering
         \includegraphics[width=\textwidth]{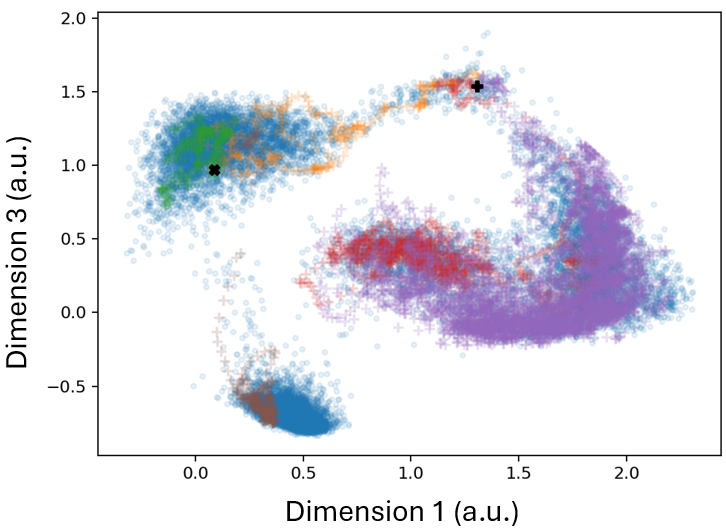}
         \caption{\label{fig:scatter_0_2}}
     \end{subfigure}
     \begin{subfigure}{0.45\textwidth}
         \centering
         \includegraphics[width=\textwidth]{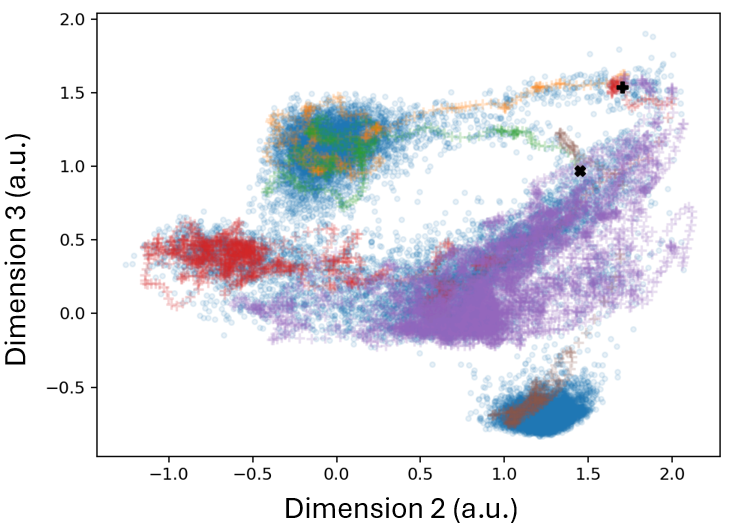}
         \caption{\label{fig:scatter_1_2}}
     \end{subfigure}
  \caption{Projection of 3D features on (a) dimensions 1 and 2, (b) dimensions 1 and 3, (c) dimensions 2 and 3, obtained after training the permutation invariant AE on all the training configurations (a subset of them are represented in blue). Additionally, the features of one (DH)I2 saddle points $\to$ \ce{2BuOH2+} trajectory (red), one (DH)I2 saddle points $\to$ \ce{iBuOH2+}-$c_2$ (orange), one (DH)I2 saddle points $\to$ secondary cation (purple), one (DH)I1 saddle points $\to$ \ce{iBuOH2+}-$c_1$ (green), one (DH)I1 saddle points $\to$ tertiary cation (brown) are plotted on top. The (DH)I1 and (DH)I2 saddle points are plotted as black "x" and "+".}
  \label{fig:scatter_ae_projections}
\end{figure}

Using the fine-tuned model, we computed the MACE atomic environment descriptors for all AIMD trajectories in the fine-tuning dataset. From the descriptors of the first 250 configurations along each trajectory, we trained the modified AE to construct three-dimensional (3D) features for a configuration from the 16 atomic environment descriptors. By visually analyzing a subset of AIMD trajectories originating from the (DH)I1 and (DH)I2 saddle points (SP), we identified the corresponding 3D features $\textbf{x}_\mathrm{enc}$ for the most relevant metastable states and transition paths. After training, the resulting latent space (or the space of the encoded features $\textbf{x}_i$ from the MACE descriptors of all atoms) clearly distinguished the two isomers \ce{iBuOH2+} and \ce{2-BuOH2+}, as well as the secondary and tertiary carbocations, as shown in Figure~\ref{fig:scatter_ae_projections}. Conversely, the two conformers of \ce{iBuOH2+} were not distinctly separated (green and orange point clouds in Figure~\ref{fig:scatter_ae_projections}). This observation justified defining a single state (basin) for \ce{iBuOH2+}, consistent with the fact that the conformational change rate is significantly faster than the isomerization and dehydration rates.

Finally, using the encoded features $\textbf{x}_\mathrm{enc}$ of four sets of configurations corresponding to \ce{iBuOH2+}, \ce{2-BuOH2+}, secondary, and tertiary carbocations, we defined a state variable $z_\mathrm{state}$ (see Equation~\eqref{eq:z_pcv} and Figure S.4b) to identify whether the system resides in one of the identified metastable states during the subsequent AMS runs. By concatenating a trajectory $\mathrm{(DH)I1} \to \ce{iBuOH2+}$ with a trajectory $\mathrm{(DH)I1} \to \mathrm{secondary}$, we constructed a progress variable $s_{i\mathrm{BuOH}_2^+ \to 2\mathrm{BuOH}_2^+}$ (see Equation~\eqref{eq:s_pcv}) to track the progress of replicas in AMS along the corresponding transition path. Further details regarding the exact parameters used to define these PCVs are provided in SI Section~5. By analyzing the values of this progress variable, $s_{i\mathrm{BuOH}_2^+ \to 2\mathrm{BuOH}_2^+}$, along trajectories initiated from the (DH)I1-SP configuration, we observed that progress along this reaction path appears to be qualitatively well indexed by this variable (see Figure S.4a), even though it was defined for the alternative path. This observation may be attributed to shared features between the paths (e.g., an increase in the C--O bond length), as reflected in the initial increase of the second dimension of the encoded features, as shown in Figures~\ref{fig:scatter_0_1} and~\ref{fig:scatter_1_2}.

When sampling initial conditions for AMS using $z_{i\mathrm{BuOH}_2^+}$, we observed rapid equilibration between the two conformers, as multiple forward and backward transitions ($c_1 \rightleftharpoons c_2$) were observed. This suggests that some initial conditions are closer to the "(DH)I1 path," while others are closer to the "(DH)I2 path." Given that $s_{i\mathrm{BuOH}_2^+ \to 2\mathrm{BuOH}_2^+}$ appears to effectively index both paths, it is likely that AMS simulations using this CV as the reaction coordinate $\xi$ and initial conditions sampled via $z_{i\mathrm{BuOH}_2^+}$ will capture both reaction pathways. This is indeed consistent with our observations. In Table~\ref{tab:isobutanol_results}, we present the estimated rate constants for this system, as well as the ratio of reaction rate constants for each product formed.

\begin{table}[!htp]
\centering
    \caption{Reaction rate constants starting from the protonated isobutanol and corresponding ratios along with the corresponding 95\% confidence interval (with $N_\mathrm{rep}=800$ and $M_\mathrm{real}=40$, a more detailed table containing the impact of $M_\mathrm{real}$ is presented in SI Section~6)}
    \label{tab:isobutanol_results}
    \begin{tabular}{|ccc|}
    \hline
    reaction & $\frac{k_\mathrm{reaction}^{Hill}}{\sum k_\mathrm{reaction}^{Hill}}$ (\%) & $k_\mathrm{reaction}^\mathrm{Hill}$ $(\mathrm{s}^{-1})$  \\
    \hline
    \ce{iBuOH2+} $\to$ \ce{2BuOH2+} &  $12.6 \pm 5.1$ & $(1.92 \pm 0.62) \times 10^{9}$ \\
    \ce{iBuOH2+} $\to$ secondary & $61.9 \pm 22.8$ & $(9.36 \pm 2.62) \times 10^{9}$\\
    \ce{iBuOH2+} $\to$ tertiary & $25.5 \pm 14.4$ & $(3.86 \pm 1.99) \times 10^{9}$ \\
    \hline
    \multicolumn{2}{|c}{reaction} &  $k_\mathrm{reaction}^\mathrm{TST}$ \\
    \hline
    \multicolumn{2}{|c}{\ce{iBuOH2+} $\to$ \ce{2BuOH2+} / secondary} & $(7.47 \pm 0.70) \times 10^{9}$\\ 
    \multicolumn{2}{|c}{\ce{iBuOH2+} $\to$  tertiary } & $(4.00 \pm 0.38) \times 10^{9}$\\ 
    \hline
    \end{tabular}
\end{table}

First of all, the estimated rates are consistent with the TST rates published in Refs.~\citenum{Gešvandtnerová2024, Gešvandtnerová2025} in the sense that estimated rate with the Hill relation are smaller than the published TST rates. Moreover, the ratio of the two rates \ce{iBuOH2+} $\to$ secondary vs \ce{iBuOH2+} $\to$ \ce{2BuOH2+} were evaluated to be 0.2 in Ref. \citenum{Gešvandtnerová2024}, which agrees well with Table~\ref{tab:isobutanol_results} where this ratio is estimated in the interval $0.20 \pm 0.09$ at a 95\% confidence. More generally, our analysis provides the following quantitative selectivity ordering among the 4 possible intermediates formed starting from  \ce{iBuOH2+}: secondary carbocation (predominant) $>$ tertiary carbocation $>$ \ce{2BuOH2+} $>>>$  \ce{tBuOH2+} (never observed). 

\section{Conclusion and Perspectives}
\label{sec:conclusion}

In this perspective, we attempted to underline the high interest of going beyond transition state theory to quantify rate constant for catalytic event thanks to the use of unbiased molecular dynamics. We showed the successful application of machine learning interatomic potentials (MLIPs) combined with the Hill relation and Adaptive Multilevel Splitting (AMS) methodology to compute reaction rate constants in two simplified but relevant systems for catalysis: water formation on $\gamma$-alumina and protonated isobutanol dehydration/isomerization in gas phase. For both examples, MLIP models fine-tuned on domain-specific datasets provided rate constants with accuracy comparable to density functional theory (DFT) calculations, but with significantly reduced computational costs. Furthermore, the combination of MLIPs and AMS enabled efficient way for sampling the reactive trajectories through molecular dynamics approach. These results highlight the potential of integrating MLIPs with rare event sampling techniques for investigating catalytic reaction mechanisms via path sampling approaches, particularly for systems where traditional DFT-based methods are computationally prohibitive.

Looking forward, the methodologies presented in this perspective hold promise for tackling more complex catalytic processes, such as the dehydration of isobutanol on alumina surfaces or within zeolite frameworks. These systems present additional challenges, including heterogeneous environments, multiple active sites, and intricate reaction networks with numerous intermediates and competing pathways. By leveraging the computational efficiency and accuracy of MLIPs alongside AMS, it becomes feasible to investigate these more complex systems with a new perspective.

From a broader point of view, path-sampling methodologies, such as AMS, offer unique advantages for studying catalytic systems. Unlike TST, which relies on strong approximations and can suffer from limitations in cases of significant anharmonicity or complex transition mechanisms, path-sampling approaches intrinsically account for dynamical effects, including recrossings and post-transition state bifurcations. While this work focused on AMS, alternative methods such as Transition Interface Sampling (TIS)\cite{vanErp2003} and Forward Flux Sampling (FFS)\cite{allen2005} exist and may lead to similar results. The development of advanced collective variable construction methods, including unsupervised and supervised machine learning approaches, further enhances the applicability of path sampling to catalytic systems. These advancements, combined with the computational power of MLIPs, may enable the exploration of realistic catalytic environments, bridging the gap between theoretical and experimental rate constant evaluation. Moreover, path sampling methods yielding samples of the reactive path ensemble, they could be used to define average reaction path using AE allowing an additional tool to analyze reaction paths.\cite{Lelièvre_2024} As a result, path-sampling approaches may play a more and more important role in future catalysis mechanisms reaction studies. 


\bibliographystyle{elsarticle-num}
\bibliography{bibli_ams_mlff_alumina}






\end{document}


\begin{frontmatter}



\title{Unbiased molecular dynamics for the direct determination of catalytic reaction times : paving the way beyond transition state theory \\ Supplementary Information} 


\author[IFPEN Solaize]{Thomas Pigeon\corref{cor1}} 
\ead{thomas.pigeon@ifpen.fr}
\author[IFPEN Solaize]{Manuel Corral Valero}
\author[IFPEN Solaize]{Pascal Raybaud\corref{cor2}}
\ead{raybaud@ifpen.fr}

\affiliation[IFPEN Solaize]{organization={IFP Energies Nouvelles},
            addressline={Rond-Point de l’Echangeur de Solaize, BP 3}, 
            city={Solaize},
            postcode={69360}, 
            state={},
            country={France}}

\end{frontmatter}



\newpage 
\section{AMS iterative procedure}

\begin{figure}[!ht]
  \centering
         \includegraphics[width=0.7\textwidth]{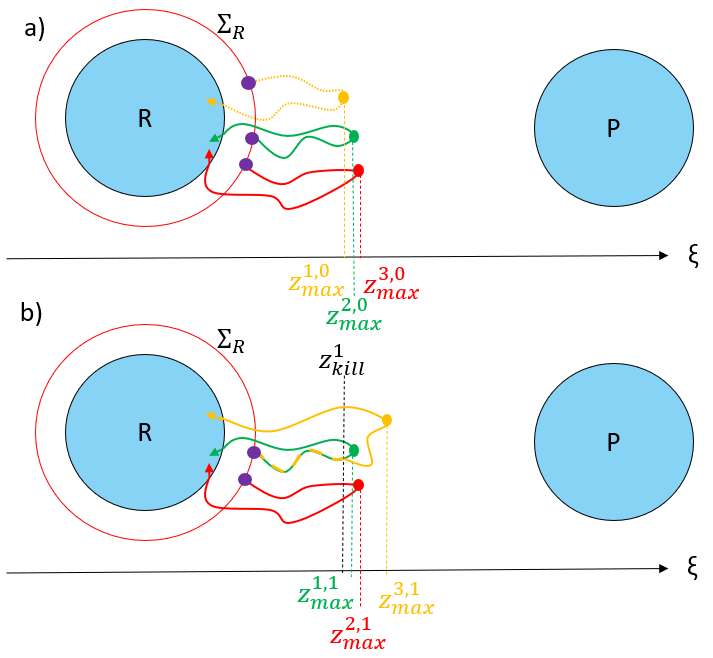}
  \caption{Schematic illustration of AMS iteration taken from Ref.~\citenum{Pigeon2023} a) The replicas are ordered by increasing maximal $\xi$ value $z_{\mathrm{max}}$ a the one corresponding to lowest $z_{\mathrm{max}} = z_\mathrm{kill}$ is killed (dashed line in yellow), b) one of the remaining trajectories is branched at the $z_\mathrm{kill}$ level and continues by running the dynamics until $R$ or $P$ is reached.}
  \label{fig:AMS_loop}
\end{figure}

In this section we present some more details concerning the iterative procedure of AMS. Each iteration $j$ can be summarized as the following sequence of actions:
\begin{enumerate}
    \item compute the maximal reaction coordinate ($\xi$) value $z_\mathrm{max}^{i,j}$ of for all trajectories;
    \item reorder replicas by increasing $z_\mathrm{max}$;
    \item identify the killing level $z_\mathrm{kill}^{j+1}$ as the $k^\mathrm{th}$ $z_\mathrm{max}$;
    \item count the number of replicas $\eta_j^{\mathrm{killed}}$ such that $z_\mathrm{max}^{i,j} \leqslant z_\mathrm{kill}^{j+1}$;
    \item regenerate $\eta_j^{\mathrm{killed}}$ replicas by branching a randomly chosen "alive" replica at the level $z_\mathrm{kill}^{j+1}$.
\end{enumerate}
An illustration of this branching procedure is also presented in Figure~\ref{fig:AMS_loop}.

\newpage 
\section{MACE-MPA-0 finetuning settings}

For both exampled studied, the fine tuning of the MACE-MPA-0\cite{batatia2023foundation} model was done a learning rate of $10^{-4}$, with a batch size of 50. The early stopping patience was set to 1. We used the exponential moving average with a decay of 0.9999. All the remaining parameters of the training script of the MACE model were kept to their default value.   

\newpage 
\section{Permutation invariant auto-encoder architecture}

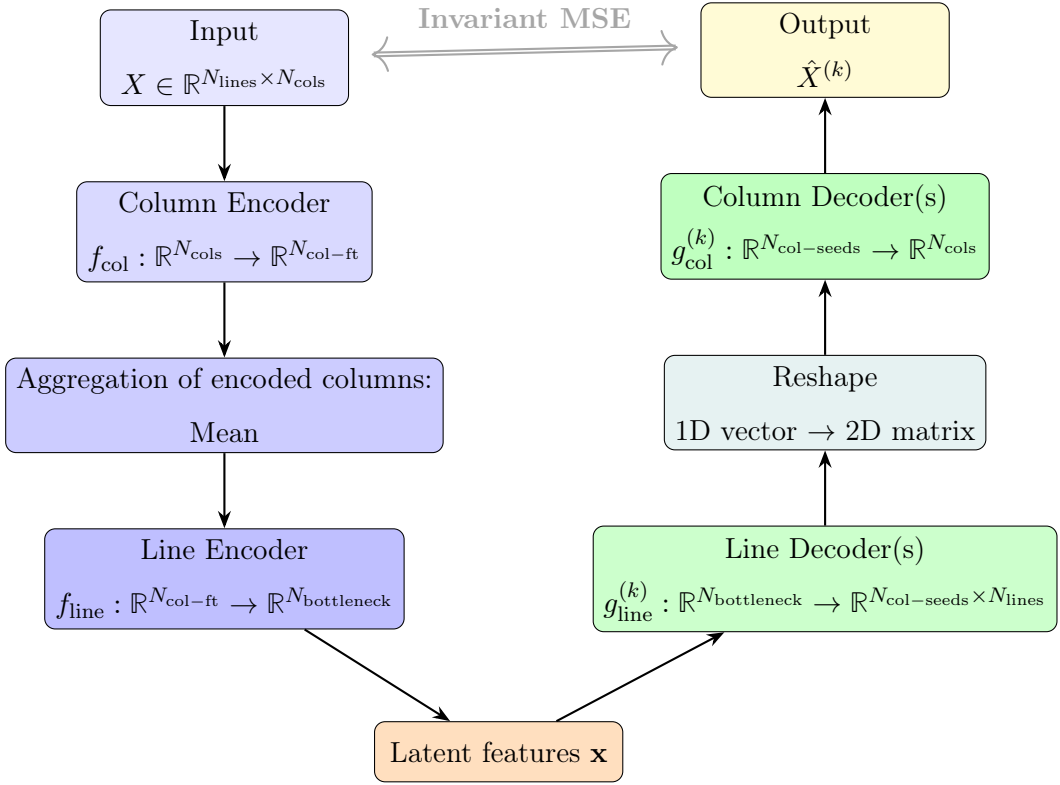
\begin{figure}[htbp]
\centering
\begin{tikzpicture}[
    node distance=10mm and 8mm,
    every node/.style={font=\small},
    box/.style={rectangle, draw, rounded corners, align=center, minimum width=3.3cm, minimum height=8mm},
    arrow/.style={-Stealth, thick}
]

\node[box, fill=blue!10] (input) {Input \\ $X \in \mathbb{R}^{N_{\mathrm{lines}}\times N_{\mathrm{cols}}}$};

\node[box, fill=blue!15, below=of input] (colenc) {Column Encoder \\ $f_{\mathrm{col}}: \mathbb{R}^{N_{\mathrm{cols}}} \to \mathbb{R}^{N_{\mathrm{col-ft}}}$};

\node[box, fill=blue!20, below=of colenc] (agg) {Aggregation of encoded columns: \\ Mean};

\node[box, fill=blue!25, below=of agg] (lineenc) {Line Encoder \\ $f_{\mathrm{line}}: \mathbb{R}^{N_{\mathrm{col-ft}}} \to \mathbb{R}^{N_{\mathrm{bottleneck}}}$};

\node[box, fill=green!20, right=2.5cm of lineenc] (linedec) {Line Decoder(s) \\ $g_{\mathrm{line}}^{(k)} : \mathbb{R}^{N_{\mathrm{bottleneck}}} \to \mathbb{R}^{N_{\mathrm{col-seeds}} \times N_\mathrm{lines}}$};

\node[box, fill=teal!10, above=of linedec] (reshape) {Reshape \\ 1D vector $\to$ 2D matrix};

\node[box, fill=green!25, above=of reshape] (coldec) {Column Decoder(s) \\ $g_{\mathrm{col}}^{(k)}:\mathbb{R}^{N_{\mathrm{col-seeds}}} \to \mathbb{R}^{N_{\mathrm{cols}}} $};

\node[box, fill=yellow!20, above=of coldec] (output) {Output \\ $\hat{X}^{(k)}$};

\coordinate (mid) at ($(lineenc.east)!0.5!(linedec.west)$);
\node[box, fill=orange!25] (bottleneck) at ($(mid)+(0,-2.3cm)$) {Latent features $\textbf{x}$};

\draw[arrow] (input) -- (colenc);
\draw[arrow] (colenc) -- (agg);
\draw[arrow] (agg) -- (lineenc);
\draw[arrow] (lineenc) -- (bottleneck);
\draw[arrow] (bottleneck) -- (linedec);
\draw[arrow] (linedec) -- (reshape);
\draw[arrow] (reshape) -- (coldec);
\draw[arrow] (coldec) -- (output);

\draw[<->, double, thick, gray!70]
  ($(input.north east)+(0.3,-0.6)$)
  -- node[midway, above, yshift=1.5mm, font=\small\bfseries, sloped=false] {Invariant MSE}
  ($(output.north west)+(-0.3,-0.6)$);
  
\end{tikzpicture}

\caption{
Schematic of the Line Permutation Invariant Autoencoder. The input matrix $x$ is first encoded column-wise by $f_{\mathrm{col}}$, aggregated with as a mean, and then encoded line-wise by $f_{\mathrm{line}}$ to produce the invariant latent features $z$. Each line decoder $g_{\mathrm{line}}^{(k)}$ expands $z$ into line-level representations, which is reshaped and finally passed to the column decoders $g_{\mathrm{col}}^{(k)}$ to reconstruct $\hat{x}^{(k)}$. The reconstruction quality is measured via an invariant mean squared error (Invariant MSE) across permutations and decoder branches.
}
\label{fig:ae_archi}
\end{figure}

The autoencoder is designed to process matrix data with enforced line permutation invariance. The architecture is composed of two main components: the encoders and the decoders, each of them being composed of column and line encoder or decoder functions. The encoders are responsible for compressing the input data into a compact latent representation, enforcing the invariance mentioned above while the decoders reconstruct the input data from this representation. The implementation of multiple decoders allows to perform a preamble clustering of the input space if desired. The architecture is illustrated in Figure~\ref{fig:ae_archi}. 

The encoding process begins with two encoders: a line encoder and a column encoder. These encoders are implemented using sequential layers consisting of fully connected layers, dropout layers for regularization, and hyperbolic tangent activation functions. The column encoder operates on the row of the input matrix and progressively reduces their dimensionality $N_\mathrm{cols}$ to a dimension $N_{\mathrm{col-ft}}$, which serves as an intermediate latent representation of the data. These resulting representations are aggregated using a mean operation over the $N_\mathrm{lines}$ to enforce invariance with respect to line permutations. The line encoder then performs a dimensionality reduction operation $f_{\mathrm{line}}: \mathbb{R}^{N_{\mathrm{col-ft}}} \to \mathbb{R}^{N_{\mathrm{bottleneck}}}$ to obtain the final bottleneck vector.

On the decoding side, the model incorporates a two-step process involving line decoders and column decoders. The bottleneck vector, obtained from the encoding stage, is first passed through the line decoder. This decoder transforms the low-dimensional latent vector into a higher-dimensional vector using a series of fully connected layers, dropout regularization, and activation functions $g_{\mathrm{line}}^{(k)} : \mathbb{R}^{N_{\mathrm{bottleneck}}} \to \mathbb{R}^{N_{\mathrm{col-seeds}}\times N_\mathrm{lines}}$. The output of the line decoder is then reshaped into a two-dimensional matrix whose number of line matches that of the original input matrix. The reshaped matrix is subsequently processed column by column using the column decoder $g_{\mathrm{col}}^{(k)}:\mathbb{R}^{N_{\mathrm{col-seeds}}} \to \mathbb{R}^{N_{\mathrm{cols}}}$. The column decoder refines the line-wise reconstructions by operating on individual columns to ensure that the final reconstruction matches the structure and dimensionality of the original input matrix.

\newpage 
\section{Permutation invariant auto-encoder training procedure}

The training process of the permutation-invariant autoencoder is designed to minimize the reconstruction error between the input and the output while handling the inherent permutation ambiguity in the data. This is achieved by introducing a mask-based mechanism that ensures the computed reconstruction error is invariant to the permutations of the input matrix's rows or columns. Below, we describe the training function, including the computation of masks and the reconstruction error.

The training begins with the forward pass, where the input data matrix $x$ is encoded into a low-dimensional bottleneck representation through the line and column encoders. This latent representation is then decoded back into a matrix of the same shape $\hat{x}^{(k)}$ as the input using the line and column decoders, as explained in the architecture description. The reconstructed matrix is compared with the original input matrix to compute the reconstruction error. However, due to the permutation-invariant nature of the task, a simple element-wise comparison is insufficient. Instead, we compute masks that align the reconstructed matrix with the input matrix in a way that minimizes the reconstruction error.

\begin{figure}[htbp]
\centering
\begin{tikzpicture}[
    node distance=9mm and 8mm,
    every node/.style={font=\small},
    box/.style={rectangle, draw, rounded corners, align=center, minimum width=4.0cm, minimum height=8mm},
    arrow/.style={-Stealth, thick}
]

\node[box, fill=blue!10] (x) {Input lines \\ $x \in \mathbb{R}^{N_{\mathrm{lines}}\times N_{\mathrm{cols}}}$};
\node[box, fill=green!25, right=5.0cm of x] (xhat) {Reconstructions \\ $\hat{x}^{(k)} \in \mathbb{R}^{N_{\mathrm{lines}}\times N_{\mathrm{cols}}}$};

\node[box, fill=orange!15, below=of $(x)!0.5!(xhat)$] (D) {Pairwise distances \\ $D_{ij}^{(k)}=\|x_i-\hat{x}_j^{(k)}\|^2$};

\node[box, fill=teal!10, below=of D] (greedy) {
  Sequential assignment, \\ see Algorithm~\ref{algo:sequential_masking}
};

\node[box, fill=orange!25, below=of greedy] (masked) {Masked sum \\ $E^{(k)}=\sum_{i,j} M_{ij} D_{ij}^{(k)}$};

\node[box, fill=green!15, below=of masked] (mse_k) {Mean per-line \\ $\mathrm{MSE}^{(k)} = \tfrac{1}{N}E^{(k)}$};

\node[box, fill=red!15, below=of mse_k] (softmin) {Softmin aggregation across decoders \\ $\mathrm{MSE}_{\mathrm{invar}}=\sum_k w_k \mathrm{MSE}^{(k)}$};

\draw (x) -- node[above,font=\scriptsize]{compare} (xhat);
\draw[arrow] ($(x)!0.5!(xhat)$) -- (D);
\draw[arrow] (D) -- (greedy);
\draw[arrow] (greedy) -- (masked);
\draw[arrow] (masked) -- (mse_k);
\draw[arrow] (mse_k) -- (softmin);

\end{tikzpicture}
\caption{
\textbf{Permutation-invariant reconstruction via sequential masking.} 
Compute the full pairwise distance matrix $D \in \mathbb{R}^{N_{\mathrm{lines}}\times N_{\mathrm{lines}}}$. Compute the binary mask $M$ via Algorithm~\ref{algo:sequential_masking} to evaluate the masked error $E^{(k)}$ which is then averaged over lines to produce $\mathrm{MSE}^{(k)}$. Multiple decoders are combined with a softmin-weighted average to produce the final $\mathrm{MSE}_\mathrm{invar}$.
}
\label{fig:ae_mse_error}
\end{figure}
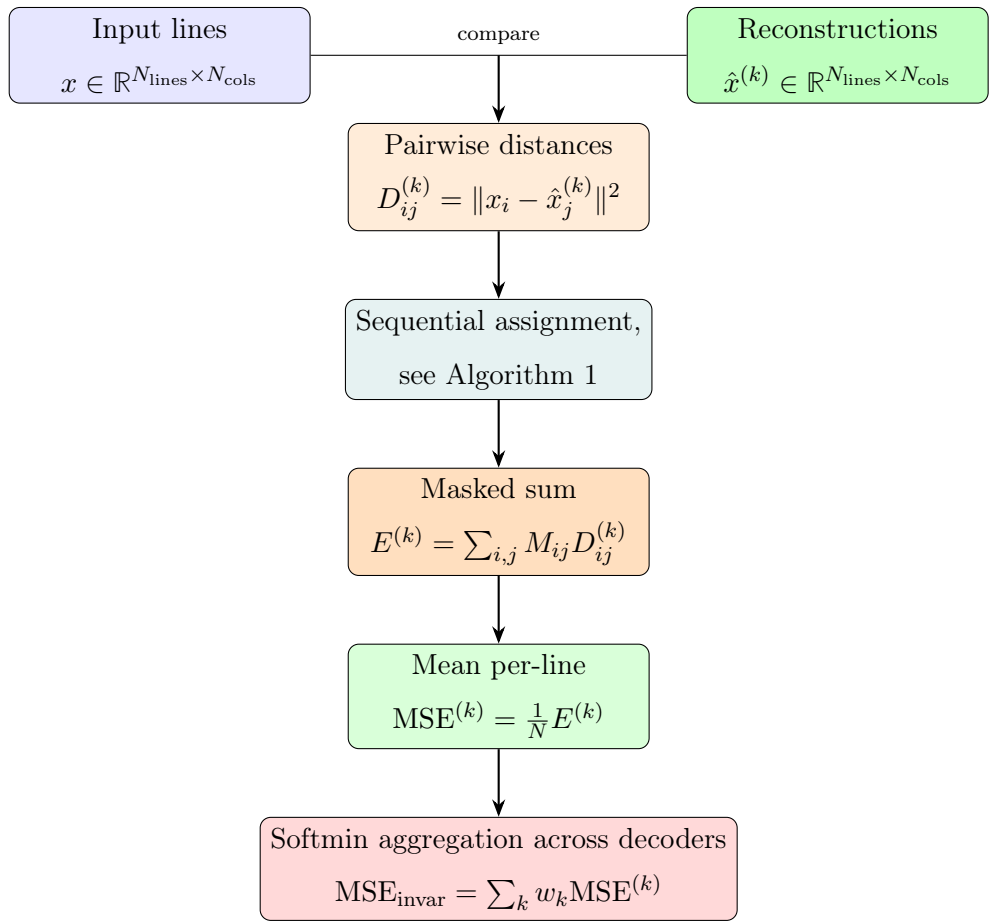

The algorithm to compute these masks focuses on finding the optimal matching between the rows of the input and reconstructed matrices. This is achieved by first defining for each decoder a pairwise distance matrix $D^{(k)}$ of size ($N_\mathrm{lines}, N_\mathrm{lines}$) whose entries $D_{i,j}$ correspond to the $L^2$ norm of the vector $x_i-\hat{x}_j^{(k)} \in \mathbb{R}^{N_\mathrm{cols}}$. Then, lines per lines, the optimal reconstruction is found sequentially, excluding the lines selected previously in the search by the procedure described in Algorithm~\ref{algo:sequential_masking}. Once the optimal matching is determined, the mask generated by Algorithm~\ref{algo:sequential_masking} is applied the matrix $D$ to compute the reconstruction error as illustrated in Figure~\ref{fig:ae_mse_error}.  This procedure is not necessarily optimal, in the sense that the mask generated this way leads to the smallest error due to the fact that it is sequential, thus a different ordering of the input may lead to lower error but it is at least cheap and seem to be sufficient to obtain satisfying results for the test cases in which we are interested in. 

\begin{algorithm}[H]
\label{algo:sequential_masking}
\SetAlgoLined
\KwIn{$D$: Distance matrix of size ($N,N$), $C$: Increment}
\KwOut{$M$: Resulting matrix}
{$M \leftarrow (0, .. 0)$} (of size ($N,N$))\; 
{$P_j \leftarrow (0, .. 0)$} (of size $N$)\; 
\For{$i \leftarrow 1$ \KwTo $N$}{
    Compute $\tilde{D}_{ij} = D_{ij} + P_j$\;
    Choose $j^\star = \underset{j}{\mathrm{argmin}} \tilde{D}_{ij}$\;
    Set $M_{i j^\star} = 1$\;
    Update $P_{j^\star} \mathrel{+}= C$\;
}
\caption{Pseudo-code for sequential masking: for each input line (in sequence), select the reconstruction (column index) minimizing a penalized distance; once a column is used it receives a large penalty $C$ so it will not be reused.}
\end{algorithm}

Finally, this error $MSE_\mathrm{invar}$ is minimized using backpropagation, with gradients computed for all parameters of the autoencoder, including the encoders and decoders. The optimization process is performed using the stochastic gradient descent-based algorithm Adam, which iteratively updates the model parameters to minimize the reconstruction error. 

\newpage 
\section{PCVs parameters}

For all PCVs, the $\lambda$ parameter was set to 2. The $z_{i\mathrm{BuOH}_2^+}$, $z_{2\mathrm{BuOH}_2^+}$, $z_{secondary}$ and $z_{tertiary}$ were constructed using 1000 configurations from a trajectory labeled by visual inspection in the corresponding metastable state. The corresponding metastable states were defined as:
\begin{equation}
\begin{aligned}
     i\mathrm{BuOH}_2^+ &: \left\{\textbf{q} \in \Omega, \, z_{i\mathrm{BuOH}_2^+}(\textbf{q}) \leqslant -6.25 \right\}  \\
     \partial \left(i\mathrm{BuOH}_2^+\right)_\varepsilon &: \left\{\textbf{q} \in \Omega, \, z_{i\mathrm{BuOH}_2^+}(\textbf{q}) = -6.20 \right\} \\ 
     2\mathrm{BuOH}_2^+ &: \left\{\textbf{q} \in \Omega, \, z_{2\mathrm{BuOH}_2^+}(\textbf{q}) \leqslant -6.5 \right\}  \\
     \mathrm{secondary} &: \left\{\textbf{q} \in \Omega, \, z_{\mathrm{secondary}}(\textbf{q}) \leqslant -6.5 \right\}  \\
     \mathrm{tertiary} &: \left\{\textbf{q} \in \Omega, \, z_{\mathrm{tertiary}}(\textbf{q}) \leqslant -6.1 \right\}  \\
\end{aligned}
\end{equation}

The $s$-PCV-\ce{iBuOH2+}$\to$\ce{2BuOH2+} is composed the 250 configurations taken from the beginnings of trajectories starting from the (DH)I2-SP configuration and respectively ending in the \ce{iBuOH2+} and \ce{2BuOH2+} meta-stable states. 

\begin{figure}[!ht]
  \centering
    \begin{subfigure}{0.472\textwidth}
    \centering
    \includegraphics[width=\textwidth]{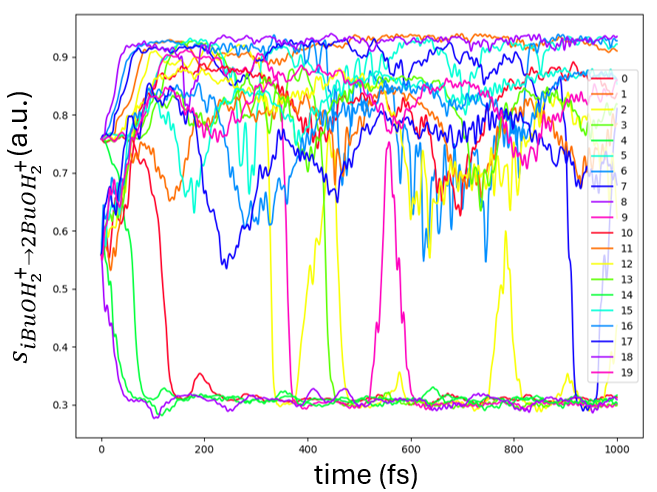}
    \caption{\label{fig:s_fig}}
    \end{subfigure}
    \begin{subfigure}{0.5\textwidth}
    \centering
    \includegraphics[width=\textwidth]{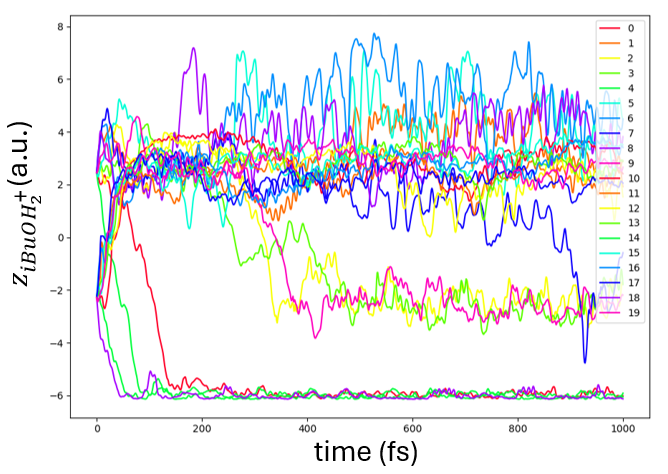}
    \caption{\label{fig:z_fig}}
    \end{subfigure}
  \caption{Plot of the evolution of $s_{i\mathrm{BuOH}_2^+ \to 2\mathrm{BuOH}_2^+}$, a) and $z_{i\mathrm{BuOH}_2^+}$ b) along 20 MD trajectories starting from (DH)I2-SP for the first 10 and (DH)I1-SP for the last 10 strucutes.}
  \label{fig:s_and_z}
\end{figure}

Figures~\ref{fig:s_and_z} illustrates the evolution of $s_{i\mathrm{BuOH}_2^+ \to 2\mathrm{BuOH}_2^+}$, and $z_{i\mathrm{BuOH}_2^+}$

\newpage 
\section{Detailed results}

\begin{table}[!htp]
\centering
    \caption{$D_1D_3 \to A_1$ reaction rate constants along with the 95\% confidence interval. $\mathrm{err ~k}$ stand for the 95\% confidence error, of the half width of the interval}
    \label{tab:alumina_results}
    \begin{tabular}{|ccccccc|}
    \hline
    entry & $N_\mathrm{rep}$ & $M_\mathrm{real}$  & $\Delta t$ $(\mathrm{fs})$& $\gamma$ $(\mathrm{s}^{-1})$ & $k_{D_1D_3 \to A_1}$ $(\mathrm{s}^{-1})$ & $\frac{\mathrm{err ~k}}{k}$ (\%)\\
    \hline
    \multicolumn{7}{|c|}{DFT} \\
    \hline
    1 & 200 & 10 & 1  & 0.5  & $(13.3 \pm 6.3) \times 10^{10}$ & 47 \\
    \hline
    \multicolumn{7}{|c|}{MACE-MPA-0 finetuned} \\
    \hline
    2 & 200 & 10 & 1    & 0.1   & $(8.72 \pm 5.66) \times 10^{10}$ & 65\\
    3 & 200 & 10 & 1    & 0.5   & $(8.93 \pm 4.49) \times 10^{10}$ & 50\\
    4 & 200 & 20 & 1    & 0.5   & $(9.23 \pm 4.27) \times 10^{10}$ & 46\\
    5 & 200 & 30 & 1    & 0.5   & $(9.61 \pm 3.64) \times 10^{10}$ & 38\\
    6 & 200 & 40 & 1    & 0.5   & $(10.6 \pm 3.07) \times 10^{10}$ & 29\\
    7 & 400 & 10 & 1    & 0.5   & $(12.3 \pm 5.55) \times 10^{10}$ & 45\\
    8 & 400 & 20 & 1    & 0.5   & $(14.2 \pm 4.10) \times 10^{10}$ & 29\\
    9 & 800 & 10 & 1    & 0.5   & $(14.3 \pm 2.57) \times 10^{10}$ & 18\\
    10 & 200 & 10 & 0.5  & 0.5   & $(10.0 \pm 3.75) \times 10^{10}$ & 37\\
    11 & 200 & 10 & 0.25 & 0.5   & $(16.8 \pm 4.00) \times 10^{10}$ & 28\\
    12 & 200 & 10 & 1    & 1     & $(12.0 \pm 6.66) \times 10^{10}$ & 55\\
    13 & 200 & 10 & 1    & 5     & $(20.3 \pm 8.68) \times 10^{10}$ & 43\\
    \hline
    \end{tabular}
\end{table}

\begin{table}[!htp]
\centering
    \caption{Reaction rate constants starting from the protonated isobutanol and selectivity along with the corresponding 95\% confidence interval}
    \label{tab:isobutanol_results}
    \begin{tabular}{|ccccc|}
    \hline
    reaction & $N_\mathrm{rep}$ & $M_\mathrm{real}$  & $k_\mathrm{reaction}$ $(\mathrm{s}^{-1})$ & $\frac{k_\mathrm{reaction}}{\sum k_\mathrm{reaction}}$ (\%)\\
    \hline
    \ce{iBuOH2+} $\to$ \ce{2BuOH2+} & 800 & 10 & $(1.51 \pm 0.60) \times 10^{9}$ & $13.0 \pm 7.0$\\
    \ce{iBuOH2+} $\to$ \ce{2BuOH2+} & 800 & 20 & $(2.29 \pm 0.93) \times 10^{9}$ & $14.1 \pm 7.4$\\
    \ce{iBuOH2+} $\to$ \ce{2BuOH2+} & 800 & 30 & $(2.08 \pm 0.76) \times 10^{9}$ & $13.8 \pm 6.5$\\
    \ce{iBuOH2+} $\to$ \ce{2BuOH2+} & 800 & 40 & $(1.92 \pm 0.62) \times 10^{9}$ & $12.6 \pm 5.1$\\
    \ce{iBuOH2+} $\to$ secondary & 800 & 10 & $(8.30 \pm 2.59) \times 10^{9}$ & $72.5 \pm 34.2$\\
    \ce{iBuOH2+} $\to$ secondary & 800 & 20 & $(10.4 \pm 3.72) \times 10^{9}$ & $63.9 \pm 31.2$\\
    \ce{iBuOH2+} $\to$ secondary & 800 & 30 & $(9.81 \pm 3.10) \times 10^{9}$ & $65.2 \pm 28.0$\\
    \ce{iBuOH2+} $\to$ secondary & 800 & 40 & $(9.36 \pm 2.62) \times 10^{9}$ & $61.9 \pm 22.8$\\
    \ce{iBuOH2+} $\to$ tertiary & 800 & 10 & $(1.67 \pm 1.63) \times 10^{9}$ & $14.4 \pm 15.0$\\
    \ce{iBuOH2+} $\to$ tertiary & 800 & 20 & $(3.60 \pm 3.12) \times 10^{9}$ & $22.1 \pm 20.5$\\
    \ce{iBuOH2+} $\to$ tertiary & 800 & 30 & $(3.15 \pm 2.17) \times 10^{9}$ & $20.9 \pm 15.6$\\
    \ce{iBuOH2+} $\to$ tertiary & 800 & 40 & $(3.86 \pm 1.99) \times 10^{9}$ & $25.5 \pm 14.4$\\
    \hline
    \end{tabular}
\end{table}

\newpage 
\bibliographystyle{elsarticle-num}
\bibliography{bibli_ams_mlff_alumina}